\documentclass[sigconf,nonacm]{acmart}
\usepackage{savesym}
\usepackage[T1]{fontenc}
\restoresymbol{TXF}{iint}
\usepackage{amsfonts}
\usepackage{amsmath}
\usepackage{textcomp}
\usepackage{bbding}
\usepackage{xspace}
\usepackage{pifont}
\usepackage[normalem]{ulem}
\usepackage{hyphenat}
\usepackage{algorithm}
\usepackage{algorithmicx}
\usepackage[noend]{algpseudocode}
\usepackage{mathtools}
\usepackage{xspace}
\usepackage{enumitem}
\usepackage{multirow}

\setlist{nosep}

\newcommand{\A}{{\tt A}\xspace}
\newcommand{\B}{{\tt B}\xspace}
\newcommand{\C}{{\tt C}\xspace}
\newcommand{\D}{{\tt D}\xspace}
\newcommand{\E}{{\tt E}\xspace}
\newcommand{\chp}{{\sf checkpoint}\xspace}
\newcommand{\pro}{{\sf propose}\xspace}
\newcommand{\acc}{{\sf accept}\xspace}
\newcommand{\cmt}{{\sf commit}\xspace}
\newcommand{\schp}{{\sf s-checkpoint}\xspace}
\newcommand{\spro}{{\sf s-propose}\xspace}
\newcommand{\sacc}{{\sf s-accept}\xspace}
\newcommand{\scmt}{{\sf s-commit}\xspace}

\colorlet{boxcolor}{teal!10!white}
\newcommand{\new}[1]{{\color{black}#1}}
\newcommand{\remove}[1]{{\color{brown}{\sout{#1}}}}

\def\sys{AdaChain\xspace}

\newcommand\vldbdoi{10.14778/3594512.3594531}
\newcommand\vldbpages{2033 - 2046}
\newcommand\vldbvolume{16}
\newcommand\vldbissue{8}
\newcommand\vldbyear{2023}
\newcommand\vldbauthors{\authors}
\newcommand\vldbtitle{\shorttitle} 
\newcommand\vldbavailabilityurl{https://github.com/chenyuanwu/AdaChain}
\newcommand\vldbpagestyle{empty} 

\newcommand{\sparagraph}[1]{\vspace{1mm}\noindent {\bf #1}\xspace}

\DeclareMathOperator*{\argmax}{arg\,max}

\newif\ifremove
\removefalse

\begin{document}
\title{\sys: A Learned Adaptive Blockchain}

\author{Chenyuan Wu}
\affiliation{%
  \institution{University of Pennsylvania}}
    \email{wucy@seas.upenn.edu}
\author{Bhavana Mehta}
\affiliation{%
  \institution{University of Pennsylvania}}
    \email{bhavanam@seas.upenn.edu}
\author{Mohammad Javad Amiri}
\affiliation{%
  \institution{University of Pennsylvania}}
    \email{mjamiri@seas.upenn.edu}
\author{Ryan Marcus}
\affiliation{%
  \institution{University of Pennsylvania}}
    \email{rcmarcus@seas.upenn.edu}
\author{Boon Thau Loo}
\affiliation{
    \institution{University of Pennsylvania}}
    \email{boonloo@seas.upenn.edu}

\begin{abstract}
This paper presents {\em \sys}, a learning-based blockchain framework that adaptively chooses
the best permissioned blockchain architecture to optimize effective throughput for dynamic transaction workloads.
\sys addresses the challenge in Blockchain-as-a-Service (BaaS) environments,
where a large variety of possible smart contracts are deployed with different workload characteristics.
\sys supports automatically adapting to an underlying, dynamically changing workload through the use of reinforcement learning.
When a promising architecture is identified, \sys switches from the current architecture to the promising one at runtime in a secure and correct manner.
Experimentally, we show that \sys can converge quickly to optimal architectures under changing workloads and
significantly outperform fixed architectures in terms of the number of successfully committed transactions, all while incurring low additional overhead.   
\end{abstract}
\maketitle

\pagestyle{\vldbpagestyle}
\begingroup\small\noindent\raggedright\textbf{PVLDB Reference Format:}\\
\vldbauthors. \vldbtitle. PVLDB, \vldbvolume(\vldbissue): \vldbpages, \vldbyear.\\
\href{https://doi.org/\vldbdoi}{doi:\vldbdoi}
\endgroup
\begingroup
\renewcommand\thefootnote{}\footnote{\noindent
This work is licensed under the Creative Commons BY-NC-ND 4.0 International License. Visit \url{https://creativecommons.org/licenses/by-nc-nd/4.0/} to view a copy of this license. For any use beyond those covered by this license, obtain permission by emailing \href{mailto:info@vldb.org}{info@vldb.org}. Copyright is held by the owner/author(s). Publication rights licensed to the VLDB Endowment. \\
\raggedright Proceedings of the VLDB Endowment, Vol. \vldbvolume, No. \vldbissue\ %
ISSN 2150-8097. \\
\href{https://doi.org/\vldbdoi}{doi:\vldbdoi} \\
}\addtocounter{footnote}{-1}\endgroup

\ifdefempty{\vldbavailabilityurl}{}{
\vspace{.3cm}
\begingroup\small\noindent\raggedright\textbf{PVLDB Artifact Availability:}\\
The source code, data, and/or other artifacts have been made available at \url{\vldbavailabilityurl}.
\endgroup
}

\section{Introduction} \label{sec:intro}

Permissioned blockchain systems
have enabled a new class of data center applications, ranging from
contact tracing~\cite{peng2021p2b}, crowdworking~\cite{amiri2021separ},
supply chain assurance~\cite{tian2017supply,amiri2022qanaat}, and federated learning~\cite{peng2021vfchain}. The popularity of these services has motivated cloud providers, e.g.,
Amazon ~\cite{amazonblockchain,qldb}, IBM~\cite{ibmblockchain}, Oracle~\cite{oracleblockchain}, and Alibaba~\cite{yang2020ledgerdb},
 to offer {\em Blockchains-as-a-Service} (BaaS)~\cite{baas2021daley}. 

BaaS offerings have resulted in a large variety of possible smart contract deployments.
Different smart contracts may exhibit different workload characteristics,
such as read/write ratios, skewness of popular keys, compute intensity, etc.
To address these variations in workloads, there has been a proliferation of permissioned blockchain systems, e.g.,
Tendermint \cite{kwon2014tendermint}, Fabric \cite{androulaki2018hyperledger}, Fabric++ \cite{sharma2019blurring},
Fabric\# \cite{ruan2020transactional}, Streamchain \cite{istvan2018streamchain}, and ParBlockchain \cite{amiri2019parblockchain}.
These systems present significant variation in architectural design, including
the number of transactions in a block,
stream processing (with no blocks), the use of reordering and early aborts, and
the sequence in which ordering, execution and validation are done.

\begin{figure}[t]
\centering
\includegraphics[width= 0.9\linewidth]{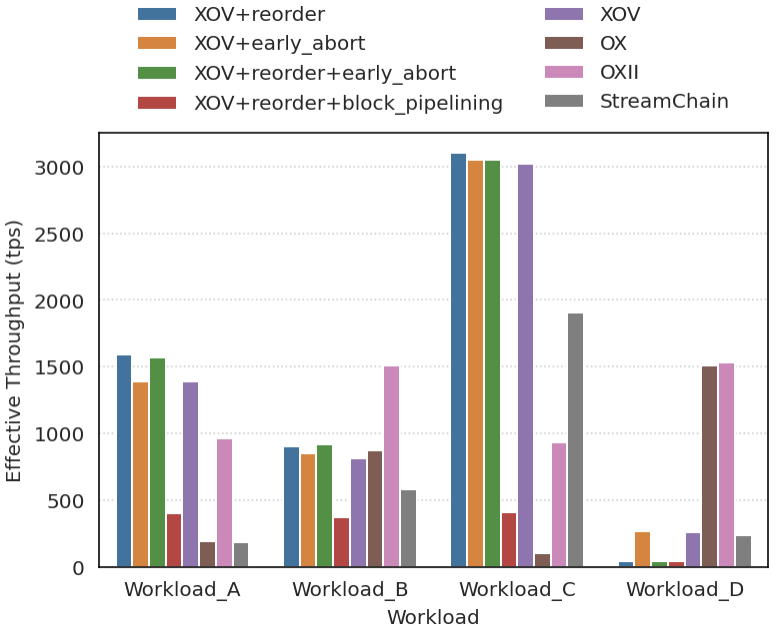}
\vspace{-1em}
\caption{Performance of different blockchain architectures under various workloads. The performance of different architectures can vary significantly between workloads. Workloads and architectures are described in Table~\ref{tbl:archs}~and~\ref{tbl:workloads}.}
\label{fig:motivation}
\end{figure}

Past studies \cite{chacko2021my,ge2022hybrid} have shown that different blockchain architectures and hyperparameter settings
are optimal for different workloads with varying properties (e.g. system load, write ratios, skewness, and compute intensity).
We experimentally confirmed this observation.
Figure~\ref{fig:motivation} shows the performance of various architectures across four different workloads, showcasing significant  variations in throughput.
For example, for Workload \A\footnote{Details about each workload can be found in Tables~\ref{tbl:workloads}~and~\ref{tbl:workload_parameters}.},
which is highly compute-intensive, an \new{Execute-Order-Validate (XOV) architecture} with reordering 
provides the best throughput.
On the other hand, for Workload \D, which requires significantly less computation but has higher skewness,
an \new{Order-Parallel Execute (OXII) architecture}
demonstrates the highest throughput.
This clearly shows the dependency between workload characteristics and the optimal blockchain architecture for each workload.

Currently, BaaS providers must choose a single architecture to offer customers,
potentially resulting in poor performance, as no single architecture provides dominant throughput. %
Even when the user has control over the blockchain architecture,
choosing the right architecture and parameters is not easy given the large configuration space. 
Moreover, in a BaaS setting, the workload may fluctuate and change, as different tenants scale up or down their smart contracts deployments,
and client requests fluctuate with different patterns throughout the day.
Of course, one could imagine building a static mapping from workload characteristics to optimal blockchain architectures --
but this mapping would (1) be expensive to compute, (2) depend on the underlying hardware, (3) still be suboptimal for workloads that shift unexpectedly over time, and
(4) require recomputing the mapping each time a new blockchain architecture is developed.

In this paper, we propose {\em \sys}, a reinforcement learning-based blockchain framework that chooses the best blockchain architecture and sets appropriate parameters in order to maximize effective throughput for dynamic transaction workloads. Experimentally, we show that \sys is not only able to select optimal or near-optimal configurations for a wide variety of workloads,
but its reinforcement learning approach also allows it to quickly adapt to new hardware,
new storage subsystems, and new unanticipated workload changes on the fly. 

In order to build an adaptive blockchain, \sys relies on two key innovations.
First, it models the selection of a blockchain architecture as a contextual multi-armed bandit problem, a well-studied reinforcement learning problem with asymptotically optimal results~\cite{thompson_bound_time}. This formulation allows \sys to apply classical algorithms, such as Thompson sampling~\cite{thompson_intro}, to select blockchain architectures in a way that minimizes \emph{regret} (the difference between the performance of the chosen architecture and the optimal architecture). \sys will strategically test different architectures to learn which ones are well-suited to the user's workload.
It learns which architectures work best by observing the characteristics of the workload and the effective throughput of the system. When the workload changes, \sys notices drops in throughput, and can automatically adjust the blockchain architecture and parameters to maximize performance, all without any user intervention.

Second, \sys introduces protocols to switch from one block\hyp{}chain architecture to another in a live system, while maintaining strong serializability properties. This switching protocol is not only required for \sys to function (multi-armed bandits generally require making multiple decisions before the optimal is reached),
but also enables a new class of blockchains that can more-or-less seamlessly transition between different architectures to support the shifting workloads in the real-world. Intuitively, the switching protocol works by splitting switching decisions between two paths. In the normal path, all nodes agree to switch to the same new architecture after a certain number of blocks have been committed, while in the slow path, all nodes switch to the same architecture after failing to make progress on processing transactions for a certain amount of time.

Specifically, this paper makes the following contributions.

\begin{itemize}[parsep=1mm, leftmargin=1em,labelwidth=*,align=left]
    \item {\bf Learned adaptive blockchain.} To the best of our knowledge,
    \sys is the first blockchain system to support
    \emph{automatically adapting to an underlying, dynamic workload}.
    Through careful modeling of the states, actions, and objective function,
    \sys's use of reinforcement learning makes it the first blockchain system to \emph{learn from its mistakes} and self-correct.
    
    \item {\bf Multi-architecture switching.} Additionally, we also present the first blockchain system
    capable of switching from one architecture to another at {\em runtime} while respecting correctness and security concerns. 
    
    \item {\bf Analysis of architectural impact on blockchain performance.}
    Through a suite of workload parameters, we perform a large-scale measurement examining the relationship
    between architecture choice and blockchain performance.
    Our experiments highlight the large state space, which renders manual heuristics difficult to achieve.
    
    \item {\bf Prototype and performance evaluation.}
    We have developed a prototype of \sys.
    Our evaluation results on CloubLab demonstrate that \sys can converge quickly to optimal architectures
    under changing workloads, significantly outperform existing fixed architectures, and incur low additional overhead.
\end{itemize}

\section{Architecture Landscape}\label{sec:arch}

\begin{table*}[t]
\caption{Comparing design principles of existing permissioned blockchain architectures. Here, P stands for performance, C stands for correctness, X stands for execution, O stands for ordering, and V stands for validation.}
\vspace{-1em}
\centering
\scriptsize
\begin{tabular}{@{}p{1cm}cccccccccc}
\toprule
Architecture & Rep. System &
P1. Block Size  & 
P2. Early Exec. &
P3. Dependency Graph &
P4. Early Abort &
P5. Cross-Block Conflicts &
P6. Parallel Exec. & 
C1. MVCC &
C2. Isolation \\ \midrule

OX  & Tendermint \cite{kwon2014tendermint} & tunable & \XSolidBrush & \XSolidBrush & \XSolidBrush  & -    & \XSolidBrush & \XSolidBrush  & strong serializable \\ 

OXII & ParBlockchain \cite{amiri2019parblockchain} & tunable & \XSolidBrush & \Checkmark & \XSolidBrush &  -    & partial & \XSolidBrush  & strong serializable \\ 

XOV & Fabric \cite{androulaki2018hyperledger} & tunable & \Checkmark & \XSolidBrush & \XSolidBrush&  V    & fully & \Checkmark  & strong serializable \\ 

XOV++ & Fabric++ \cite{sharma2019blurring} & tunable & \Checkmark & \Checkmark & X, O&  V    & fully & \Checkmark  & strong serializable \\ 

XOV\# & Fabric\# \cite{ruan2020transactional} & tunable & \Checkmark & \Checkmark & O&  O    & fully & \XSolidBrush  & serializable \\ 

XOV & StreamChain \cite{istvan2018streamchain} & 1 & \Checkmark & \XSolidBrush & \XSolidBrush &  V    & fully & \Checkmark  & strong serializable \\ 

\bottomrule
\end{tabular}
\label{tbl:archs}
\end{table*}

\begin{table}[t]
\caption{Characterizing workloads A, B, C and D. Specific workload parameters are presented in Table~\ref{tbl:workload_parameters}.}
\vspace{-1em}
\scriptsize
\begin{tabular}{cccccc}
\toprule
Workload & Write Ratio &
Contention Level  & 
Load &
Compute Intensity \\ \midrule

A  & low & high & high & high  \\ 

B & moderate & high & moderate & low \\ 

C & moderate & low & high & very high  \\ 

D & high & very high & moderate & very low \\
\bottomrule
\end{tabular}
\vspace{1em}
\label{tbl:workloads}
\end{table}

To motivate \sys, we first examine, both intuitively and experimentally, why different blockchain architectures perform diversely across different workloads.
In Section~\ref{sec:blockchain_arch_analysis}, we highlight a number of blockchain architectures, and illustrate their advantages and disadvantages. The point here is not that some blockchain architectures are always ``better'' or ``worse'' than others, but rather that each blockchain architecture performs well under some conditions and poorly under others. In Section~\ref{sec:case_for_adapt}, we argue that a blockchain that can adaptively switch between multiple architectures is able to achieve ``best of all worlds'' performance.

\vspace{-0.5em}
\subsection{Blockchain Architectures and Workloads}
\label{sec:blockchain_arch_analysis}

Table~\ref{tbl:archs} lists representative blockchain systems and their architectures, where
the design space consists of seven performance optimizations (P1-P7) and two correctness dimensions (C1-C2).
Figure~\ref{fig:motivation} shows their performance under four different workloads.
Here, we use \textit{effective} throughput as the performance metric, i.e., the number of successfully committed transactions per second. 

The workloads \A to \D are characterized in Table~\ref{tbl:workloads}. BaaS workloads embody a large extent of variations.
For instance, different transactions might invoke different percentages of write operations to the underlying key-value store, as represented by the {\em write ratio}. These transactions might also contend to access or update the same set of popular keys (or hot keys), as indicated by the {\em contention level}.
In addition, the runtime {\em load} on a BaaS can be determined by the frequency of issued transactions by each client and the number of active clients varying with time.
Last, {\em compute intensity} is an important characterization of BaaS workloads, as pointed out by~\cite{wu2022flexchain, ruan2020transactional, gramoli2022diablo}. This is because permissioned blockchains support a wide range of applications, some of which are compute-intensive (e.g., those that provide security and correctness guarantees for machine learning applications).

Below, we briefly describe the design principles of each architecture and explain the intuition behind why the performance of each architecture can vary under different workloads.

\sparagraph{Order-Execute (OX).}
The order-execute architecture has been widely used in permissioned blockchain systems such as Tendermint~\cite{kwon2014tendermint}, Quorum~\cite{morgan2016quorum},
Chain Core~\cite{chain}, Multichain~\cite{greenspan2015multichain},
Hyperledger Iroha~\cite{iroha}, and Corda~\cite{corda}.
In the OX architecture, transactions are totally ordered and batched into blocks and then transactions of a block are executed sequentially.
As a result, the OX architecture does not require a \new{Multi-version Concurrency Control (MVCC)} validation phase, which is used to resolve conflicts between transactions,
and hence, no transactions will be aborted due to conflicts.
As shown in Figure~\ref{fig:motivation}, this design principle makes OX outstanding at workload \D,
where transactions are write-heavy and contentious, i.e.,  transactions update a small set of hot keys.
On the other hand, OX performs comparatively poorly on workloads \A and \C, which are compute-intensive.
Due to the lack of parallel execution mechanisms, OX cannot take advantage of the multi-core processing power of modern servers. 

\sparagraph{Order-Parallel Execute (OXII).}
In the OXII architecture, used by ParBlockchain~\cite{amiri2019parblockchain},
transactions are first totally ordered and batched into blocks.
OXII then constructs a dependency graph for transactions within a block based on their positions.
Specifically, if $t_i$ is ordered before $t_j$, and the pair of transactions are conflicting,
OXII adds an edge from $t_i$ to $t_j$.
This dependency graph is then used to execute transactions in parallel,
i.e., a transaction can be executed once all its predecessors have finished execution.
Given a higher level of execution parallelism than OX, OXII performs better than OX under computation-heavy workloads such as \A and \C.
Note that even for a given workload, OXII requires careful tuning of block size;
a large block results in high overhead in dependency graph construction,
while a small block results in less parallelism and higher communication overhead.

\sparagraph{Execute-Order-Validate (XOV).}
Hyperledger Fabric~\cite{androulaki2018hyperledger} presents the XOV architecture 
(which was first introduced by Eve~\cite{kapritsos2012all} in the context of Byzantine fault-tolerant SMR)
by switching the order of the ordering and execution phases such that
transactions are simulated fully in parallel before being ordered in the ordering phase.
Since it utilizes early execution, XOV requires an MVCC validation phase to
invalidate all transactions that are simulated on stale data,
and commits only the validated transactions to the world-state and the blockchain ledger.
This early execution enables XOV to perform well on contention-free workloads such as \C.
On the other hand, XOV demonstrates poor performance under contentious and write-heavy workloads, such as \B and \D,
due to the high percentage of invalidated transactions.
\new{Similarly, as network delay increases, XOV suffers from inconsistent world states among peers as well as stale reads, and thus significantly degraded performance~\cite{chacko2021my}.}

\sparagraph{XOV with early abort and reordering (XOV++).}
The XOV++ architecture, as introduced in Fabric++~\cite{sharma2019blurring},
follows the XOV paradigm but with some modifications.
First, a dependency graph is constructed in the ordering phase to capture RW conflicts between each pair of transactions within the same block.
When the graph is constructed, all elementary cycles in the graph 
are aborted greedily.
Unlike OXII, which utilizes the graph for concurrency control,
XOV++ uses the graph for transaction reordering;
when there is a RW conflict between $t_i$ and $t_j$, it (re)orders $t_i$ before $t_j$ in the block.
Second, it adopts early abort techniques in both the simulation and ordering phases.
Whenever XOV++ detects that a transaction operates on stale data,
XOV++ immediately aborts that transaction without waiting for the final MVCC validation.
As an effect of transaction reordering, XOV++ has outstanding performance on workload \A,
where the conflicts are reconcilable given a low write ratio.
On the other hand, XOV++ performs poorly on workload \D with a near-zero effective throughput.
This is because, under a contentious and update-heavy workload, very few conflicts can be reconciled through reordering.
Moreover, reordering becomes more expensive when there are a large number of cycles in the dependency graph,
resulting in more pending blocks and, thus, more transactions that simulate on stale data.

\sparagraph{XOV with serializable isolation (XOV\#).}
The XOV\# architecture, presented in Fabric\#~\cite{ruan2020transactional}, is mainly different from XOV and XOV++ in that XOV\# is serializable, while XOV and XOV++ are strong serializable.
To achieve this isolation level, XOV\# incrementally constructs a dependency graph
that keeps track of all dependencies, including those that span across blocks in the ordering phase.
Once a transaction is ordered, XOV\# immediately drops this transaction if there is a dependency cycle involved.
The resulting acyclic schedule is guaranteed to be serializable, thus, no extra MVCC validation is needed in XOV\#.
To ensure a fair comparison with other architectures, we run XOV\# under the strong serializability isolation level
while keeping the remaining design dimensions the same as the original XOV\# (the XOV+reorder+block\_pipelining bar).
XOV\# performs worse than vanilla XOV in all workloads \A to \D
due to the overhead of maintaining a large dependency graph and detecting cycles.
This suggests that the performance improvement reported in Fabric\# is mainly due to a more relaxed isolation level.

\sparagraph{Stream XOV.}
StreamChain~\cite{istvan2018streamchain} switches from block processing to stream transaction processing.
Specifically, StreamChain follows the XOV paradigm while fixing the block size to $1$.
The motivation behind stream processing is simple: while the original, permissionless blockchains were forced to used proof of work (PoW) consensus techniques to maintain fault tolerance, a permissioned blockchain environment allows more efficient consensus protocols to be used.
Thus, stream processing can reduce transaction latency.
In terms of effective throughput, StreamChain has relatively good performance when the workload is lightweight or not contentious,
such as in workloads \B and \C.
Otherwise, the high block construction overhead in terms of cryptographic operations and excessive disk I/Os leads to
a large number of pending blocks in StreamChain, making incoming transactions simulate on stale data. %
\new{As expected, StreamChain is more sensitive to the type of storage hardware used than OX: the system have poor performance without RAM disk~\cite{chacko2021my}. }
StreamChain also highlights that the \emph{parameters} of a given architecture can impact performance. A large block size leads to higher block formation overhead and latency, while a small block size results in higher communication and disk overhead. 

\new{
\subsection{The Case for Reinforcement Learning}
\label{sec:case_for_adapt}

The previous subsection demonstrates that depending on workload and hardware characteristics, the performance of a given blockchain architecture can vary drastically. We thus argue that \emph{there is no one-size-fits-all architecture.} 

One may consider the design of simple heuristics to map workload characteristics and hardware features to the optimal blockchain architecture. However, designing a good heuristic is difficult and cumbersome in the case of permissioned blockchains. For example, Hyperledger Fabric~\cite{androulaki2018hyperledger} has proposed that ``when the contention level is high, use OX; otherwise, use XOV''. Unfortunately, such a simple heuristic leads to wrong decisions in $50\%$ of our recorded traces, let alone in a real BaaS production environment. 

This suggests, to develop a good heuristic, an expert needs to exhaustively experiment over the entire state and action space to understand the intricate interactions. The expert also needs to carefully tune the specific threshold parameters that distinguish ``very high contention'' from ``high contention'', or ``low write ratio'' from ``moderate write ratio'', etc. Given the large space of workload (i.e., write ratio, contention level, load, compute intensity) and hardware (i.e., CPU, RAM, disk, network) features, if the expert discretizes each dimension by uniformly sampling 5 points in it and tests each architecture for 5 seconds, it takes at least 90 days to conduct the experiments even when the effect of block size is ignored. This is a conservative estimate given that most dimensions will require more than 5 sample points in practice. Due to the lack of blockchain simulators that capture the variations in architectures,
such experiments are hard to be conducted in parallel. Moreover, such heuristics become outdated when new hardware or blockchain architectures are introduced in the BaaS and hence, difficult for the expert to keep up with these changes. For instance, when the BaaS provider introduces non-volatile RAM~\cite{tsai2020disaggregating} to replace the traditional combination of volatile RAM and disk, or when BaaS transitions to disaggregated architecture~\cite{wu2022flexchain} where there is extra latency for accessing memory, new hardware features are also introduced.

Reinforcement learning (RL) is an ideal solution to this Sisyphean task, which has shown superior performance in other learned systems~\cite{neo,bao}. Unlike supervised learning that assumes training data is complete and requires a separate data collection process prior to deployment, RL-based system learns from its mistakes and optimizes long term rewards through its trials.  With reinforcement learning, \sys can optimize itself to whatever workloads, hardware and blockchain architectures at hand, providing significant \textit{operational benefits}.

}

\section{\sys Overview} \label{sec:overview}

At a high level, \sys contains two key components:
a machine learning model (the \emph{learning agent})
which guides \sys towards better and better blockchain architectures, and
an \emph{architecture switching} mechanism that allows \sys to near-seamlessly transition from one blockchain architecture
to another while ensuring correctness and security. 

\sparagraph{Learning agent.}
\sys's learning agent models the problem of selecting a blockchain architecture as a contextual multi-armed bandit (CMAB) problem~\cite{bandit_survey}:
periodically, \sys examines the most recent properties of the workload (\emph{context}),
and then selects one of many blockchain architectures (\emph{arms}).
After making the selection, it observes the effective throughput of the newly selected architecture (\emph{reward}).
To be successful, \sys must balance the \emph{exploration} of new, untested architectures
with \emph{exploiting} past experience to maximize throughput --
without a careful balance of exploration and exploitation,
\sys risks failing to discover an optimal configuration (too much exploitation),
or performing no better than random (too much exploration).
We select this CMAB formulation (as opposed to generalized reinforcement learning models)
because CMABs are exceptionally well-studied, and many asymptotically-optimal algorithms exist to solve them~\cite{thompson_intro, thompson_bound}.
Details about the learning agent are provided in Section~\ref{sec:learningalgo}.

\sparagraph{Switching architecture.}
\sys utilizes a switching protocol that allows it to switch from one blockchain architecture
to another in a distributed fashion across all nodes in the blockchain deployment, while transactions are ongoing.
\sys achieves this by splitting switching decisions between two paths,
a normal path in which all nodes agree to switch to the same new architecture after a certain number of blocks have been committed,
and a slow path in which all nodes switch to the same new architecture after failing to make progress for a certain amount of time.
Details about \sys's switching protocol are in Section~\ref{sec:switching}.

\begin{figure}[t]
\centering
\includegraphics[width=\linewidth]{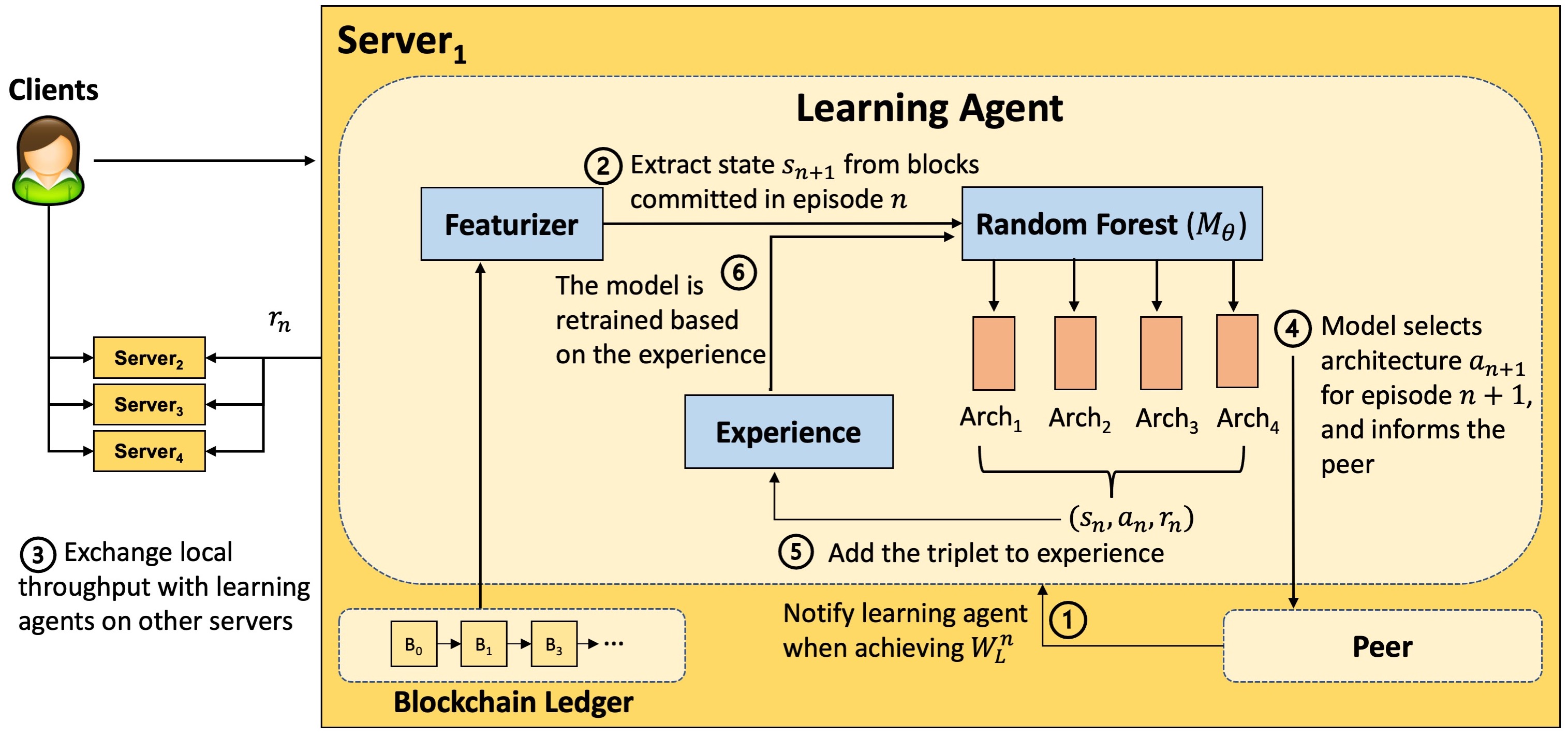}
\vspace{-1em}
\caption{Architecture of \sys. For readability, we only present the internals of $\text{server}_1$.}
\label{fig:arch}
\end{figure}

\sparagraph{\sys workflow overview.}
Figure~\ref{fig:arch} shows the overall architecture of \sys.
\sys operates in \emph{episodes}, where within one episode, the blockchain architecture remains unchanged.
When the learning agent finds an architecture candidate, it instructs the peer to use that architecture for the next episode.
Each episode is marked by the completion of a constant number of transactions ($\Delta N_{\text{episode}}$), including invalidated transactions.
This episode design ensures \sys does not stuck in a bad architecture for a long time
even when the fraction of invalidated transactions is high due to conflicts. 
Below, we describe how the learning agent proposes the architecture for episode $n+1$ in detailed steps.
Although our discussion below focuses on the internals of $\text{server}_1$,
the same procedure happens simultaneously on every blockchain server. 

\sparagraph{Step 1: Notifying the learning agent.} 
In episode $n$, the peer notifies its local learning agent when the number of committed blocks has reached a certain watermark.
The notification also includes the local performance measurement $r_n$ in episode $n$.

\sparagraph{Step 2: Featurization.}
Since learning agents are distributed across different servers,
the state (i.e., some features that capture the workload) that they need in order to make a decision should also be distributed.
In \sys, states are not only distributed, but also decentralized as no single entity controls the state.
This is possible with negligible overhead due to two key insights:
(1) the blockchain ledger contains rich information about the workload, and thus is a good source of raw data for featurization;
(2) the ledger is naturally decentralized and consistent across peers.  Thus, once the agent is notified by the peer, its featurizer extracts the state $s_{n+1}$ from blocks committed in episode $n$. Details are described in Section~\ref{sec:state}.

\sparagraph{Step 3: Exchanging performance measurements.}
The locally observed performance is different across different peers, and malicious peers could even manipulate its local measurement.
To ensure each honest server has the same architecture for episode $n+1$,
the learning agent on $\text{server}_1$ exchanges the local measurement $r_n$ with learning agents on every other server,
so as to agree on performance measurement. \new{Details are described in Section~\ref{sec:switching} and ~\ref{sec:security}.}

\sparagraph{Step 4: Estimating the performance for each architecture.}
The predictive model $M_\theta$ predicts the performance of each architecture candidate under state $s_{n+1}$,
and selects architecture $a_{n+1}$ that is predicted to have the best performance.
The learning agent then informs the peer to switch to $a_{n+1}$ for episode $n+1$. 

\sparagraph{Step 5: Building experience buffer.}
Once an $r_n$ is obtained, the learning agent adds the $(s_n, a_n, r_n)$ triplet to the experience buffer.
Note that $s_n$ and $a_n$ are derived prior to the start of episode $n$. 

\sparagraph{Step 6: Retraining.}
The predictive model $M_\theta$ is periodically retrained based on the experience buffer, creating a feedback loop.
As a result, \sys's predictive model improves, and \sys more reliably picks the best architecture for the observed state.  

\sparagraph{Assumptions.} In short, \sys can adapt itself according to the workload and hardware setup to continually improve performance.
Moreover, \sys is an \textit{online} learned system that does not require a separate and cumbersome data collection process prior to deployment. 
Our current design makes two assumptions.
First, in \sys, similar to many other permissioned blockchain systems \cite{kwon2014tendermint,amiri2021sharper,morgan2016quorum},
each node serves as both the ordering and execution (endorser) node.
This, however, is in contrast to Hyperledger Fabric and its variants that separate endorsing and ordering roles.
Second, \sys is designed for a homogeneous setup, where different servers have access to similar resources.
While having these two assumptions in place simplifies the system design, they have been used in real-world BaaS deployments.

\section{Learning Algorithms} \label{sec:learningalgo}

In this section, we discuss \sys's learning approach in detail. We first formalize \sys's learning problem as a contextual multi-armed bandit problem, and then discuss our selected algorithm, Thompson sampling, for solving such problems. We next describe the predictive model used by \sys, followed by the specific state and action space design.

\sparagraph{Contextual multi-armed bandits (CMABs).} In a contextual multi-armed bandit problem, an agent periodically makes decisions in a number of episodes, enumerated by $n$. In each episode, the agent selects an action $a_n$ based on a provided state $s_n$, and then receives a reward $r_n$. The agent's goal is to select actions in a way that minimizes \emph{regret}, i.e., the difference between the reward from the chosen action and the reward from the optimal action. CMABs assume that each episode is independent\footnote{Contextual multi-armed bandit algorithms have been shown to be effective even when these condition do not strictly hold~\cite{thompson_intro}.} from each other, and that the optimal decision depends only on the state $s_n$. As described in Section~\ref{sec:overview} and~\ref{sec:switching}, in order to be responsive to workload changes, there are no pending blocks across different episodes in \sys. Thus, each episode in \sys can also be considered to be independent (although caching effects may bring a small amount of dependence between episodes).

\sparagraph{\sys's formulation.} \label{sec:mlformulation}
\sys uses \textit{effective} throughput as the performance metric $P$ to maximize, which is the number of successfully committed transactions per second.
For each episode, it must select an architecture to use.
\sys's goal is to select the best architecture (in terms of effective throughput) in the family of available architectures $A$,
given the current perceived workload $w \in W$. We call this selection function $S: W \to A$.
We formalize the goal as a regret minimization problem, where the regret $r_n$ for an episode $n$
is defined as the difference between the effective throughput of the architecture selected by \sys and the ideally optimal architecture
as presented in equation~\ref{eq:formulation}.
\begin{equation}
r_n =  \mathop{\max}_{ a \in A} P(w, a)  - P(w, S(w))
\label{eq:formulation}
\end{equation}

\new{We use effective throughput as the performance metric since it is the dominant metric used by blockchain benchmark tools~\cite{dinh2017blockbench, hyperledgercaliper, gramoli2022diablo} and previous literatures that proposed fixed architectures~\cite{androulaki2018hyperledger, sharma2019blurring, ruan2020transactional, amiri2019parblockchain}. Extending the optimization goal to a combination of metrics is left for future work.}

\sparagraph{Thompson sampling.}
While there are many algorithms to solve contextual multi-armed bandit problems, we select Thompson sampling for its simplicity: at the start of each episode, we train a model based on our current experience, and then select the best action as predicted by the model. In order to train the model, Thompson sampling deviates from traditional ML techniques
: instead of selecting the model parameters that are most likely given the data, we \emph{sample model parameters proportionally to their likelihood given the training data}. More formally, we can define maximum likelihood estimation as finding the model parameters $\theta$ that maximize likelihood given experience $E$: $\argmax_\theta P(\theta \mid E)$ (assuming a uniform prior). Instead of maximizing likelihood, Thompson sampling simply samples from the distribution $P(\theta \mid E)$. This means that if we have a lot of data suggesting that our model weights should be in a certain part of the parameter space, our sampled parameters are likely to be in the part of the space. Conversely, if we have only a small amount of data suggesting that our model weights should be in a certain part of the parameter space, we may or may not sample parameters in that part of the space during any given episode. 

\subsection{Predictive Model} \label{sec:predictivemodel}
\sys uses random forests~\cite{rf} as the predictive model due to their good performance on data sets of moderate sizes and fast inference. The model takes the state (i.e., workload) concatenated with action (i.e., architecture choice) as input, and outputs the predicted performance.\footnote{This corresponds to a \emph{value based model}. A \emph{policy} model, in which the predictive model predicts simultaneously the probability of each action being optimal, might be an interesting direction for future work.} Thus, given a state, \sys enumerates the action space and uses the model to predict the performance of each action. \sys then chooses the action with the best-predicted performance to be carried out. Once there is a tie on the best-predicted performance, \sys breaks the tie randomly to avoid local maxima. %

Integrating random forests with Thompson sampling requires the ability to sample model parameters from $P(\theta\mid E)$. The simplest technique (which has been shown to work well in practice~\cite{thompson_bootstrap}) is to train the model as usual, but only on a bootstrap~\cite{bootstrapping} of the training data. In other words, the random forest is trained using $|E|$ random samples drawn with replacement from experience $E$, inducing the desired sampling properties. \sys uses this bootstrapping technique for its simplicity. 

\subsection{State Space} \label{sec:state}
In \sys, the state represents properties of the client workload. \sys captures the state space using the four features below.
To ensure the accuracy of feature extraction, all aborted or invalidated transactions are still written to the ledger with a validity flag
(similar to~\cite{fabric2018collections}).
Below, we assume a window of blocks $b_i$ to $b_j$ in the ledger are read by the learning agent for featurizing the current state.

\noindent \textit{\uline{Write ratio.}} We observe that counting the write ratio in terms of write accesses to the key-value store is not effective for predicting performance. Thus, \sys measures the write ratio at the transaction level: once a transaction writes to the key-value store, it is viewed as a write transaction. The write ratio is the ratio of write transactions to the number of all transactions during $b_i$ and $b_j$.

\noindent \textit{\uline{Hot key ratio.}} \sys measures the frequency that each key is accessed during $b_i$ and $b_j$. It then takes the frequency corresponding to the hottest key to be the hot key ratio.

\noindent \textit{\uline{Transaction arrival rate.}} \sys timestamps each transaction upon its first arrival to the system.
\sys first measures the number of all transactions from $b_i$ to $b_j$, denoted as $N$,
and then derives the transaction arrival rate using $\frac{N}{ts_j - ts_i}$, where
$ts_i$ represents the arrival timestamp of the first transaction in $b_i$ and
$ts_j$ represents that of the last transaction in $b_j$.

\noindent \textit{\uline{Execution delay.}} \sys uses average execution delay of all transactions in the period of $b_i$ to $b_j$.

\subsection{Action Space} \label{sec:action}
In \sys, the action space consists of different blockchain architectures.
One na\"{\i}ve approach to represent the action space is to give each architecture a one-hot encoding.
However, from the random forest's perspective, this approach makes two semantically close architectures totally unrelated,
resulting in ineffective splits, and thus poor prediction accuracy.
For example, assume XOV is represented by vector $(1, 0, ..., 0)$ in the one-hot encoding.
Random forest might split on the first dimension in the vector,
i.e., XOV is its left child, while everything non-XOV is its right child.
Each child's performance will be predicted using the average performance of that child.
Clearly, XOV++ and StreamChain might have a relatively close performance to XOV,
but they will always fall into a wrong child node and their predicted performances are wrongly averaged.

Thus, \sys chooses to first featurize the blockchain architectures to maintain the semantic information of their design. Feature engineering an optimal representation of blockchain architectures is a difficult and inexact task. Instead of attempting to design an all-encompassing representation that captures every dimension of blockchain architectures, we instead selected a simple representation based on our intuition of the most important properties. We leave investigating alternative representations to future work.

\sys therefore captures the action space using three main features:
block size, early (speculative) execution, and dependency graph construction. Block size is a scalar variable, representing the number of transactions within a block.
The block size in \sys is also equal to the batch size in the consensus protocol.
To limit the growth of action space, the block size can not exceed $1,000$,
which is larger than typical block sizes used in blockchain systems,
and we further discretize the block size using paces. Early execution and dependency graph construction are both binary variables. Thus, \sys's action space consists of $100$ choices in total.

We do not consider parallel execution as a feature
because it can be derived from the two previous features (i.e., early execution and dependency graph construction): (1) early execution of transactions happens fully in parallel; (2) the goal of constructing a dependency graph is to execute independent transactions in parallel.

\section{Switching Architectures} \label{sec:switching}

This section discusses the architecture switching mechanism of \sys.
We first introduce the normal path of operations, followed by our timeout-based mechanism in the slow path.

\newcommand{\CHP}{\footnotesize \textsf{CHECKPOINT}\xspace}
\newcommand{\SCHP}{\footnotesize \textsf{S-CHECKPOINT}\xspace}
\newcommand{\PRO}{\footnotesize \textsf{PROPOSE}\xspace}
\newcommand{\SPRO}{\footnotesize \textsf{S-PROPOSE}\xspace}
\newcommand{\ACC}{\footnotesize \textsf{ACCEPT}\xspace}
\newcommand{\SACC}{\footnotesize \textsf{S-ACCEPT}\xspace}
\newcommand{\CMT}{\footnotesize \textsf{COMMIT}\xspace}
\newcommand{\SCMT}{\footnotesize \textsf{S-COMMIT}\xspace}
\newcommand{\SWCH}{\footnotesize \textsf{SWITCH?}\xspace}

\begin{algorithm}[t]
\footnotesize
\caption{Normal path}\label{alg:regularpath}
\begin{algorithmic}[1]

\Statex {\color{teal} $\rhd$ On each server $i$}
\State \textbf{Upon} index of local last committed block $b_{\text{last}}$ reaching $W_L^n$
\State \quad Record performance $p_i^n$
\State \quad Extract features $f_i^{n+1} = (w_i^{n+1}, c_i^{n+1}, r_i^{n+1}, e_i^{n+1})$ from block $W_H^{n-1}$ to $W_L^n$
\State \quad Multicast  $\langle \CHP, n, i, e_i^{n+1}, p_i^n\rangle_{\sigma_i}$ to all servers
\Statex {\color{teal} $\rhd$ On the leader sever $l$}
\State \textbf{Upon receiving} valid \CHP messages from a quorum {\sf Q} of $2f+1$ servers
\State \quad Compute $e^{n+1} \gets \text{median} \{e_j^{n+1} | j \in$ {\sf Q}$\}$
\State \quad Compute $p^{n} \gets \text{median} \{p_j^n | j \in$ {\sf Q}$\}$
\State \quad Multicast $\langle\langle \PRO, e^{n+1}, p^n \rangle_{\sigma_l}, \mathcal{C} \rangle$ to all servers
\Statex {\color{teal} $\rhd$ On each server $i$}
\State  \textbf{Upon receiving} a \PRO message from the leader
\State \quad \textbf{if} $e^{n+1}$ and $p^n$ are valid (based on $\mathcal{C}$) \textbf{then}
\State \qquad Multicast $\langle \ACC, n, i, e^{n+1}, p^n \rangle_{\sigma_i}$ to all servers
\State \textbf{Upon receiving} valid matching \ACC messages from $2f+1$ different servers
\State \quad Multicast $\langle \CMT, n, i, e^{n+1}, p^n \rangle_{\sigma_i}$ to all servers
\State \textbf{Upon receiving} valid matching \CMT messages from $2f+1$ different servers
\State \quad Add $p^n$ to experience and derive action $a_{n+1}$ based on $f^{n+1}$

\State \textbf{if} $T^n$ transactions have been committed \textbf{then}
\State \quad Abort any new incoming transaction $t$ in the ordering phase

\State \textbf{Upon} $b_{\text{last}}$ reaching $W_H^n$
\State \quad Pause block formation thread until action $a_{n+1}$ is derived
\State \quad $W_L^{n+1} \gets W_H^n + \lfloor  \Delta N_{\text{learn}} / |b_{n+1}| \rfloor$
\State \quad $W_H^{n+1} \gets W_H^n + \lfloor  \Delta N_{\text{episode}} / |b_{n+1}| \rfloor$
\State \quad $T^{n+1} \gets T^n + \lfloor  \Delta N_{\text{episode}} / |b_{n+1}| \rfloor \times |b_{n+1}|$
\State \quad $n \gets n+1$
\State \quad Carry out action $a_{n+1}$
\State \quad Reset timer $\tau$

\end{algorithmic}
\end{algorithm}

\subsection{Normal Path} \label{sec:regularpath}

Algorithm~\ref{alg:regularpath} presents the normal path of operations.
Each server in \sys runs Algorithm~\ref{alg:regularpath} in a distributed fashion
in order to carry out architecture switching.
Here, $S$ is the set of blockchain servers, $i$ stands for the index of the server, $n$ stands for the current episode, and
$\Delta N_{\text{episode}}$ and $\Delta N_{\text{learn}}$ are two constant hyperparameters.
At a high level, the normal path introduces two watermarks:
a low watermark ($W_L^n$) that triggers the learning phase, and
a high watermark ($W_H^n$) that marks the end of an episode. 

The untrustworthiness of participants in a blockchain system prevents us from relying on
a centralized entity to featurize the state and measure the reward.
Thus, inspired by the PBFT protocol~\cite{castro1999practical}, \sys conducts them in a decentralized fashion.
Upon reaching the low watermark $W_L^n$, each server $i \in S$ records its locally observed throughput $p_i^n$ of episode $n$
and featurizes the state for the next episode $n+1$ from its local blockchain ledger (lines 1-3).
Although most dimensions of the state are naturally consistent across different servers,
there can be slight variations on the execution delay, $e_i^{n+1}$, and measured throughput.
Thus, each server $i$ multicasts a \chp message consisting of $e_i^{n+1}$ and $p_i^n$ to all other servers (line 4).
\sys relies on the leader server to
(1) collect a quorum {\sf Q} of $2f+1$ \chp messages,
(2) calculate the median of observed throughput values to be the global reward $p^n$, and
(3) calculate the median of the execution delay values $e^{n+1}$ to be part of the global state (lines 5-7).
Once both values are computed, the leader multicasts a \pro message, including the values and
the set $\mathcal{C}$ of $2f+1$ received \chp messages to all servers (line 8). 
Upon receiving the \pro message, each server validates the message according to the set $\mathcal{C}$ and multicasts an \acc message
to all other servers (lines 9-11).
Each server then waits for $2f+1$ matching \acc messages before sending a \cmt message (lines 12-13).
The \acc and \cmt phases, similar to {\sf prepare} and \cmt phases of PBFT,
ensure that values are correct and replicated on a sufficient number of nodes.
Finally, when a server receives $2f+1$ \cmt messages,
the predictive model will derive action $a_{n+1}$ as described in Section~\ref{sec:learningalgo} (lines 14-15).
Note that since \acc and \cmt messages are broadcast to all servers,
even if a server has not received the \pro message from the leader
(due to the asynchronous nature of the network or the maliciousness of the leader),
the server still has access to the values.

In order to be responsive to workload changes, each episode is marked by the completion of
$\lfloor  \Delta N_{\text{episode}} / |b_{n}| \rfloor$ blocks, where 
$\Delta N_{\text{episode}}$ is a constant hyperparameter of the system
($10,000$ transactions in the current deployment) and $|b_{n}|$ denotes the block size in episode $n$.
As a result, each episode processes $\lfloor  \Delta N_{\text{episode}} / |b_{n}| \rfloor \times |b_{n}|$ transactions,
including transactions invalidated in MVCC validation due to conflicts.
Specifically, when the number of committed transactions \textit{in consensus} reached $T^n$,
\sys early aborts transactions in the ordering phase (i.e., no more transactions will be committed by the consensus protocol)
until \sys transitions into the next episode (lines 16-17).
In \sys, the block formation thread waits for transactions to be committed,
cuts the block, possibly performs dependency graph construction, reordering, or execution according to the current architecture,
and lastly, commits the block.
Once the number of committed blocks reaches the high watermark,
the block formation thread will be paused until action $a_{n+1}$ is derived (lines 18-19).
This ensures exactly $\lfloor  \Delta N_{\text{episode}} / |b_{n}| \rfloor$ blocks
are committed in episode $n$ on different servers.
Note that the learning phase (including feature extraction, exchanging measurements, training, and inference)
is triggered by low watermark $W_L^n$, and \sys keeps processing transactions using architecture $a_n$ between $W_L^n$ and $W_H^n$.
Thus, as shown in Section~\ref{sec:expoverhead}, architecture $a_{n+1}$ is derived before reaching $W_H^n$ in most cases,
ensuring high throughput of the system.

\begin{algorithm}[t]
\footnotesize
\caption{Slow path}\label{alg:slowpath}
\begin{algorithmic}[1]
\Statex {\color{teal} $\rhd$ On each server $i$}
\State \textbf{if} timer $\tau$ timeouts and $b_{\text{last}}$ has not reached $W_L^n$ \textbf{then}
\State \quad pause block formation thread after committing the current block
\State \quad Record performance $p_i^n$
\State \quad Multicast  $\langle \SCHP, n, i, b_{\text{last}, i}^n\rangle_{\sigma_i}$ to all servers

\Statex {\color{teal} $\rhd$ On each server $j$ where $\tau$ has not been expired}
\State \textbf{Upon receiving} $f+1$ valid \SCHP messages from different servers
\State \quad pause block formation thread after committing the current block
\State \quad Record performance $p_j^n$
\State \quad Multicast  $\langle \SCHP, n, j, b_{\text{last}, j}^n\rangle_{\sigma_j}$ to all servers
\Statex {\color{teal} $\rhd$ On the leader sever $l$}
\State \textbf{Upon receiving} valid \SCHP messages from a quorum {\sf Q} of $2f+1$ servers
\State \quad Compute $W_H^n \gets \text{max} \{b_{\text{last}, j}^n | j \in$ {\sf Q}$\}$
\State \quad Multicast $\langle\langle \SPRO, W_H^n \rangle_{\sigma_l}, \mathcal{C'} \rangle$ to all servers
\Statex {\color{teal} $\rhd$ On each server $i$}
\State  \textbf{Upon receiving} a \SPRO message from the leader
\State \quad \textbf{if} $W_H^n$ is valid (based on $\mathcal{C'}$) \textbf{then}
\State \qquad Multicast $\langle \SACC, n, i, W_H^n \rangle_{\sigma_i}$ to all servers

\State \textbf{Upon receiving} valid matching \SACC from $2f+1$ different servers
\State \quad Multicast $\langle \SCMT, n, i, W_H^n \rangle_{\sigma_i}$ to all servers
\State \textbf{Upon receiving} valid matching \SCMT from $2f+1$ different servers
\State \quad Resume block formation thread
\State \quad Extract features $f_i^{n+1} = (w_i^{n+1}, c_i^{n+1}, r_i^{n+1}, e_i^{n+1})$ from block $W_H^{n-1}$ to $b_{\text{last}}$
\State \quad Multicast  $\langle \CHP, n, i, f_i^{n+1}, p_i^n\rangle_{\sigma_i}$ to all servers
\Statex {\color{teal} $\rhd$ On the leader sever $l$}
\State \textbf{for} every transaction $t$ in the ordering phase \textbf{do}
\State \quad $t.episode \gets n$
\Statex {\color{teal} $\rhd$ On each server $i$}
\State \textbf{for} every transaction $t$ committed by consensus \textbf{do}
\State \quad \textbf{if} $t.episode \neq n$ \textbf{then}
\State \qquad abort $t$

\end{algorithmic}
\end{algorithm}

\subsection{Slow Path} \label{sec:slowpath}

Before \sys converges to the optimal architecture, the learning agent might occasionally choose ``bad'' architectures.
The bad architectures might result in a high fraction of transactions being invalidated, or a slow growth of committed blocks (e.g., choosing OX when the workload is highly compute-intensive,
or choosing XOV+reorder when the contention is extremely high).
In terms of wall-clock time, \sys should not be stuck in either scenario.
While the normal path is capable of handling the first scenario,
we further introduce a slow path to handle the scenario where the growth of committed blocks is slow.

Algorithm~\ref{alg:slowpath} presents the slow path operations.
When server $i$ timeouts and the index of the last committed block, $b_{\text{last}}$, has not reached the low watermark,
server $i$ pauses the block formation after committing the current block,
records the performance $p_i^n$  in the current episode,
and multicasts a \schp message including the $b_{\text{last}}$ to all servers (lines 1-4).
If a server $j$ receives \schp messages from at least $f+1$ servers, even if its timer has not expired,
it pauses the block formation, records its performance, and multicasts a \schp message to all servers (lines 5-8).

When the leader receives \schp messages from a quorum of $2f+1$ servers,
it finds the maximum index of the last committed block across all servers, $W_H^n$, and
multicasts a \spro message including $W_H^n$ and the received $2f+1$ \schp messages to all servers (lines 9-11).
All servers validate the received \spro message before two rounds of \sacc and \scmt communication, as shown in lines 12-16
(similar to the normal path).
Each server then uses $W_H^n$ as its high watermark and then resumes the block formation thread (lines 17-18). 
This ensures that in a slow path, the same number of blocks are committed in episode $n$ across different servers.
These operations might be expensive on the normal path, but are negligible on the slow path,
compared to the timeout ($15$s in our case) and the poor performance before timeouts.
The worst case happens when a fast server has not sent a \schp message,
or its message has not been considered in the leader's calculation of $W_H^n$.
In this case, if the index of its last committed block is higher than $W_H^n$, the server needs to rollback those exceeding blocks.
Similar to the normal path, each server also needs to exchange state and performance measurements
to derive action $a_{n+1}$ for the next episode (lines 19-20).

Upon receiving transaction $t$ for ordering at the leader $l$,
the leader tags $t$ with the current episode $n$ as part of the sequence number (lines 21-22).
When a server receives transactions committed by consensus protocol,
it aborts transactions whose tagged episode is not equal to the current episode (lines 23-25).
This ensures episode independence, i.e., there are no pending blocks across episodes in \sys.
As a result, a bad architecture that triggers the slow path
will not affect the performance of future episodes with promising architectures.

The normal path and slow path of \sys ensure two properties.
First, transactions are strong serializable, and
second, the world state is eventually consistent across different servers.

\new{\section{Security Analysis} \label{sec:security}

Compared to fixed-architecture blockchains, our use of machine learning and run-time architecture switching add new security risks. In this section, we briefly present the new threats and discuss how \sys addresses them. 

\sparagraph{Adversarial ML.} As studied in the ML community, machine learning can be adversarial~\cite{goodfellow2014explaining, kurakin2016adversarial, madry2018towards, wong2019fast}. In the context of \sys , using reinforcement learning does not affect the \emph{correctness} of the system, which depends only on the consensus protocol (predefined in our system), the current architecture, and our switching protocol. That being said, \sys's correctness guarantee is always as strong as the weakest architecture in its action space, regardless of the specific reinforcement learning algorithm.

However, reinforcement learning introduces a new \emph{performance} attack vector: manipulating feature data to cause the learning agent to pick a bad architecture. To carry out this attack, a malicious node might propose adversarial feature values into the quorum $Q$ in Algorithm~\ref{alg:regularpath}. There are at least two ways such adversarial features could negatively impact performance: (1) decision attacks that target the inference phase, where an adversary reports false observations of its own features in order to push the global feature in one direction or another; and (2) poisoning attacks that target the training phase~\cite{poison}, where an adversary reports carefully selected feature values and labels to cause the next trained model to be inaccurate.
\sys mitigates such attacks by selecting the median value of all reported features. For a feature with low variance, adversaries would only be able to move the median value by a small amount even if they can report strong outliers (e.g., infinity or zero). However, for a feature with high variance, adversaries could potentially create a non-trivial change in the median value, impacting future inference and training. Since \sys is designed to operate in a homogeneous environment, most existing features and labels have low variance.

Thus, while adversarial ML attacks cannot impact correctness, there are open questions about potential performance attacks caused by learning. Studying the full impact of adversarial learning on a system like \sys is an interesting avenue for future work.

\sparagraph{Choosing different architectures.} If peers choose different architectures within the same episode, the world state across peers can diverge and lead to correctness issues. \sys guarantees that every honest node agrees on the same new architecture in the same episode. We provide an analysis as follows. 

Each node's learning agent starts with the same random seed when it is launched. Thus, since both the state and reward of a certain episode are the same across all honest nodes (as mentioned in Section~\ref{sec:switching}), with the predictive model's deterministic training and inference, each honest agent chooses the same blockchain architecture in the same episode. Moreover, although dishonest nodes are tempted to make decisions that are different from honest nodes, they cannot affect the agreement on architecture among honest nodes. Without loss of generality, we assume $f=1$ and there are $3f+1$ peers in the system, where $P_1, P_2, P_3$ are honest and $P_4$ is malicious; we further assume the honest peers choose the XOV architecture. 

\noindent {\em \uline {Case 1: $P_4$ is the leader in the consensus protocol.}} There are four possible scenarios, none of which poses a correctness issue: (1)
if $P_4$ chooses OX in the same episode, it will send out transaction proposals instead of endorsements in the ordering phase, so honest peers will detect this mismatch and initiate a view-change; (2) if $P_4$ chooses XOV but with a batch size $b_4$ that is different from what honest learning agents have chosen, it does not affect the agreement; (3) if $P_4$ chooses XOV with different batch sizes and uses different batch sizes for different peers (e.g., $b_{4, 1}$ for $P_1$, $b_{4, 2}$ for $P_2$, etc.), the honest peers will detect and reject these batches during consensus on batches and initiate a view-change; and (4) finally, if $P_4$ chooses XOV but with an opposite reordering choice, since reordering happens locally on each peer according to its local model, the malicious peer cannot corrupt honest peers. 

\noindent {\em \uline {Case 2: $P_4$ is not the leader in the consensus protocol.}} The honest leader detects type mismatch for messages originating from $P_4$ and discards them, while other honest peers work as normal.

\sparagraph{Delay in architecture switching.} 
An adversary might deliberately delay its own communication
(sending messages) during architecture switching. In this scenario, if the adversary is a backup node in the switching protocol, it does not hurt the system’s throughput assuming the number of faulty backups is less than $f$. On the other hand, if the adversary is the leader, it could carry out a performance attack by delaying the transition into the new architecture, where the delay is carefully chosen to avoid triggering the timeout. Fortunately, there are known orthogonal techniques to mitigate these attacks. For example, instead of using a pessimistic switching protocol inspired by PBFT, \sys can adopt a robust switching protocol, e.g., following Prime~\cite{amir2011prime} which is a BFT consensus protocol robust to such timeout attacks.

\sparagraph{Other vanilla threats.} Other common Byzantine failures might also occur during the normal path and slow path operations. For instance, a malicious leader could send different \pro messages to different backups, or forge a deviated global state and reward. In the meantime, a malicious backup node could double vote. \sys handles these threats using a PBFT-style switching protocol, which guarantees that the committed state and reward of a certain episode is the same across all honest nodes. We refer readers to the original paper~\cite{castro1999practical} for more details.
}
\section{Evaluation}\label{sec:eval}

Our evaluation aims to answer the following questions: 
\begin{enumerate}[parsep=1mm, leftmargin=0em,labelwidth=*,align=left]
\item  Can \sys converge to the optimal architecture under a static workload without prior knowledge? (Section~\ref{sec:expstatic})
\item  How well does \sys perform compared to existing fixed blockchain architectures when the workload changes? (Section~\ref{sec:expchanging})
\item  How does the hardware setup (e.g., the type of CPU, network latency and bandwidth, etc.) affect the performance of \sys and existing blockchain architectures? (Section~\ref{sec:exphardware})
\item What overhead does \sys introduce? (Section~\ref{sec:expoverhead})
\end{enumerate}

\subsection{Experimental Setup}\label{sec:expsetup}
We have implemented a prototype of \sys in C++ and Python.
The blockchain peers which process transactions and carry out architecture switching are implemented in C++.
We use gRPC for communications between peers and LevelDB~\cite{leveldb} for storing the world states.
The learning agents are implemented separately in Python due to its mature machine learning libraries.
Each peer communicates with its local learning agent through gRPC.

\noindent \textbf{Testbed.}
Our testbed consists of 4 c6220 bare-metal machines on CloudLab~\cite{duplyakin2019design},
each with two Xeon E5-2650v2 processors ($8$ cores each, $2.6$Ghz), 64GB RAM(8 x 8GB DDR-3 RDIMMs, 1.86Ghz)
and two 1TB SATA 3.5'' 7.2K rpm hard drives.
These machines are connected by two networks, each with one interface:
(1) a 1 Gbps Ethernet control network;
(2) a 10 Gbps Ethernet commodity fabric.
Unless otherwise specified, we use the second network for all communication.
We set the size of the execution thread pool equal to the number of cores on each peer.

\noindent \textbf{System configuration.}
We run a single blockchain channel consisting of 3 peers on 3 different servers.
As mentioned in Section~\ref{sec:overview}, each peer in \sys serves as an executor as well as an orderer.
The choice of consensus protocol is configurable inside \sys, and we use~Raft \cite{ongaro2014search} for consistency with Hyperledger Fabric and its variants.
We run the client on a separate server with 3 threads, each firing transaction proposals to one specific peer.
The reported throughput only considers \textit{effective} transactions, i.e., excluding early aborted and invalidated transactions.
Throughout this paper, we parameterize the architecture switching protocol as follows:
normal path timeout is set to $15$s, the low watermark is set to $7500$ transactions, and the high watermark to set to $10000$ transactions.

\noindent \textbf{Workloads.}
To capture the diversity in real-world blockchain transactions,
we implement a benchmark driver above SmallBank~\cite{dinh2017blockbench}
to derive customized workloads with tunable parameters.
The benchmark driver preloads the blockchain with 10k users, each with two accounts.
We set $N_{hot}$ of them as hot accounts.
When firing transactions, the client randomly picks one of the five modifying transactions with probability $P_w$
and the read-only transactions with probability $1 - P_w$.
Each transaction has a certain probability to access the hot accounts, as controlled by the $P_{hot}$ parameter.
The client continuously fires $N_{trans}$ transactions every $T_{fire}$ milliseconds.
To simulate computation-heavy transactions, each transaction has a $T_{compute}$ interval
after it fetches the required world states from the key-value store and before its subsequent operations. 

\begin{table}
\caption{Specific workload parameters.}
\scriptsize
\begin{tabular}{crrrrrrr}
\toprule
Workload & $P_w$ & $P_{hot}$  & $N_{hot}$ & $N_{trans}$ & $T_{fire}$ & $T_{compute}$ \\ \midrule
A  & 0.2 & 0.95 & 5 & 300 & 50ms & 5ms   \\ 
B & 0.5 & 0.99 & 10 & 100 & 50ms & 1ms \\ 
C & 0.5 & 0.1 & 10 & 300 & 50ms & 10ms \\ 
D & 0.9 & 0.95 & 1 & 100 & 50ms & 0ms \\ 
E & 0.5 & 0.99 & 10 & 100 & 50ms & 5ms \\
\bottomrule
\end{tabular}

\vspace{0.5em}
\label{tbl:workload_parameters}
\end{table}

We use workloads \A-\E throughout this paper, where workloads \A-\D are the same as in Figure~\ref{fig:motivation}.
The specific parameters of workloads \A-\E are listed in Table~\ref{tbl:workload_parameters}.
Unlike workloads \A-\D, the additional workload \E is introduced later in the section that explores adaptivity to different hardware settings.
Although workload \E has only slight deviation from workload \B, it renders blockchain architectures extremely sensitive to hardware setup (details in Section~\ref{sec:exphardware}).
Note that we have written our own benchmark driver because
no existing benchmark captures all these variations in workloads.

\begin{figure*}
\begin{minipage}{0.245\textwidth}
	\centering
	\includegraphics[width=\columnwidth]{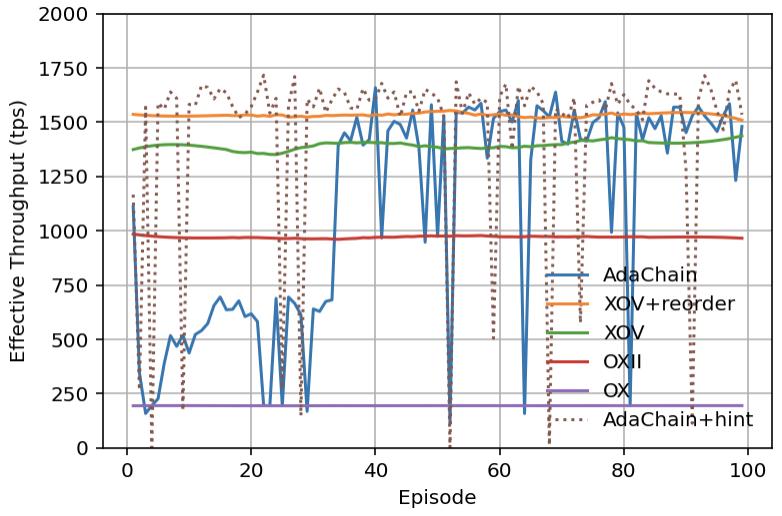}
    \vspace{-2.2em}
    \captionsetup{format=plain, font=scriptsize}
    \caption*{\normalfont \,\,\,\,\, (a) Workload A}
\end{minipage}
\begin{minipage}{0.245\textwidth}
	\centering
	\includegraphics[width=\columnwidth]{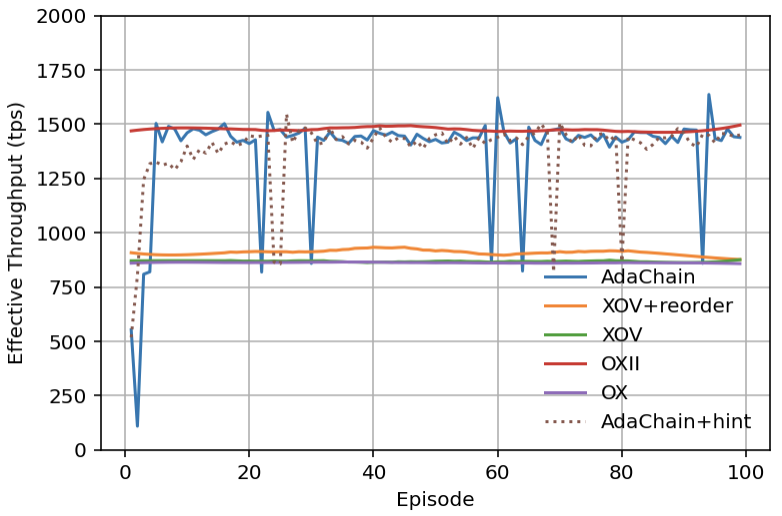}
    \vspace{-2.2em}
    \captionsetup{format=plain, font=scriptsize}
    \caption*{\normalfont \,\,\,\,\, (b) Workload B}
\end{minipage}
\begin{minipage}{0.245\textwidth}
	\centering
	\includegraphics[width=\columnwidth]{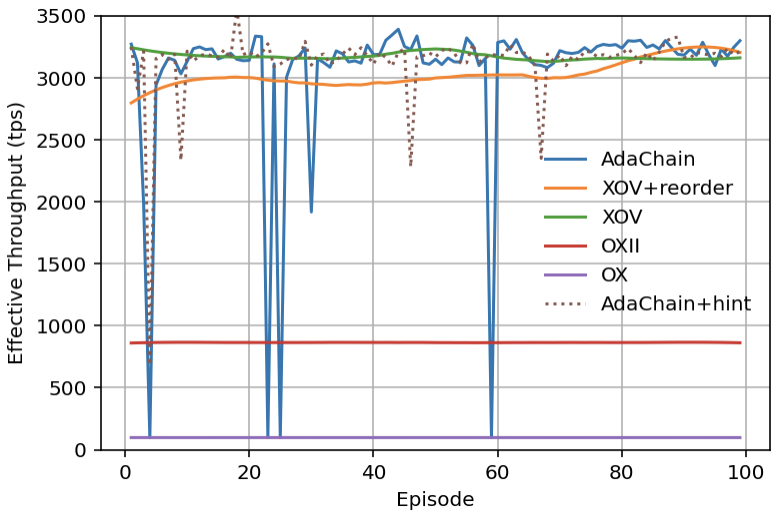}
    \vspace{-2.2em}
    \captionsetup{format=plain, font=scriptsize}
    \caption*{\normalfont \,\,\,\,\, (c) Workload C}
\end{minipage}
\begin{minipage}{0.245\textwidth}
	\centering
	\includegraphics[width=\columnwidth]{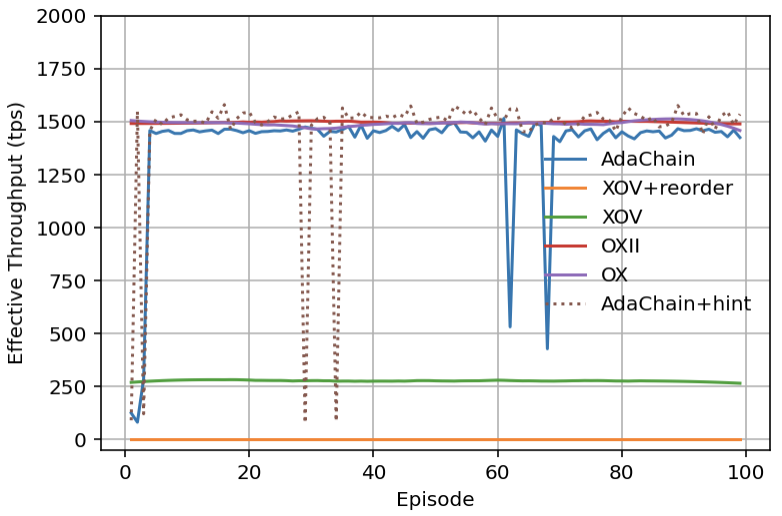}
    \vspace{-2.2em}
    \captionsetup{format=plain, font=scriptsize}
    \caption*{\normalfont \,\,\,\,\, (d) Workload D}
\end{minipage}
\vspace{-1em}
\captionsetup{labelfont={color=black, bf}}
\caption{\color{black} Convergence of \sys to the optimal architecture under static workloads. Each plot shows the performance of \sys (w/ and w/o hint), XOV+reorder, XOV, OXII, and OX. While different blockchain architectures are optimal for each workload, \sys always reaches near-optimal performance.}
\label{fig:static}
\end{figure*}

\vspace{-1em}
\subsection{Convergence under Static Workloads}\label{sec:expstatic}
Our first set of experiments aims to demonstrate that \textit{\sys can rapidly converge to the optimal architecture under a static workload with no prior experiences required}. We run \sys for 100 episodes on four representative workloads (i.e., \A-\D).
To compare \sys against the \emph{a priori} optimal architecture,
we also perform a grid search in the action space to find the optimal architecture for each workload.
For our four workloads, we compare \sys with the four optimal static architectures. 

Figure~\ref{fig:static} plots the performance of \sys and four baselines for each workload,
where the baseline curves are smoothed 
for better readability.
The curve for \sys is not smoothed.
Each workload has a different optimal architecture.
For instance, XOV+reorder is optimal for workload \A, suboptimal for workloads \B and \C, but the worst for workload \D.
Unlike a fixed architecture that cannot adjust itself even under a static workload,
\sys always converges to the optimal architecture for each workload quickly within $40$ episodes,
no matter how bad the first episode (initial architecture) is.
Due to Thompson sampling, \sys still performs some exploration in the architecture space even after convergence,
as identified by the drops in the performance plot.
Although these explorations do not seem useful under static workloads,
they are crucial for finding optimal architectures under a changing workload which is more realistic in today's BaaS environment. 

Comparing to exhaustive grid search (not shown in Figure~\ref{fig:static}) that takes $n_a$
(the size of action space, i.e., $100$ in our case) to converge and performs pure random exploration at all times,
\sys converges much faster and strikes a better balance between exploiting known good actions and
exploring unknown actions. \new{Table~\ref{tbl:avg_tp} shows \sys's typical convergence time. In our definition, convergence is reached when staying within $95\%$ of the optimal performance for $5$ consecutive episodes.}

\begin{table}
\caption{Effective throughput (tps) for each architecture in the last $20$ episodes of each workload and
\color{black}the convergence time (minutes) of \sys.}
\scriptsize
\begin{tabular}{c|rrrrr|rr}
\toprule
\multirow{2}{*}{Workload} & \multicolumn{5}{c|}{Effective Throughput}  & \multicolumn{2}{c}{\new{\sys's Conv. Time}} \\ 
& XOV+reorder & XOV & OXII & OX & \sys & \new{w/ hint} & \new{w/o hint} \\
\midrule
A        &\textbf{1532}  &  1415         &   968        &  194         &  1425  &       \new{0.65}&  \new{2.48} \\
B        &     897       &  866          &\textbf{1545} &  861         &  1426  &       \new{0.42} & \new{0.62}  \\
C        &     3228      & \textbf{3235} &   940        &   98         &  3153  &       \new{0.45}  & \new{0.48}\\
D        &     1         &  272          &  1494        & \textbf{1498}&  1447  &       \new{0.43}   &\new{0.43}\\ 
\midrule
Average  &     1414      & 1447          &  1237        &  663         &  \textbf{1862} & \new{0.49}  & \new{1.00}\\
Worst   &     1          & 272           & 940          & 98           & \textbf{1425} &   \new{0.65}  &\new{2.48} \\
\bottomrule
\end{tabular}

\vspace{0.5em}
\label{tbl:avg_tp}
\end{table}

\sparagraph{Average performance.} \sys obviously does not outperform the optimal action in any workload (it is the optimal action, after all).
However, we show that \sys does offer good average and worst-case performance after convergence.
Table~\ref{tbl:avg_tp} shows the throughput for each blockchain on each workload in the last $20$ episodes of execution.
Even with a few performance drops due to exploration, \sys achieves both the best average throughput \emph{and}
the best worst-case throughput across all four workloads.

\new{\sparagraph{Providing hints.} An experienced administrator can build his knowledge into the learning framework by specifying certain rules in order to prevent some sub-optimal decisions. An example hint is ``if the compute intensity is higher than $2000$ us, enable early execution; otherwise, disable early execution". If accurate, hints of this nature reduce the search space, thus providing faster convergence and avoid certain explorations. Figure~\ref{fig:static} summarizes our findings using the example hint (see ``AdaChain+hint'' lines). For workload \A, the hint accelerates the convergence time of \sys by $3.8\times$. However, it is not sufficient to avoid all explorations in \sys,  i.e., varying reordering choice and block size. In contrast, for workloads \B and \C where the convergence is already fast, the additional hint reduces unnecessary explorations after convergence. }

\vspace{-1em}
\subsection{Adaptivity under a Changing Workload}\label{sec:expchanging}

\begin{figure}
\centering
\includegraphics[width=0.7\linewidth]{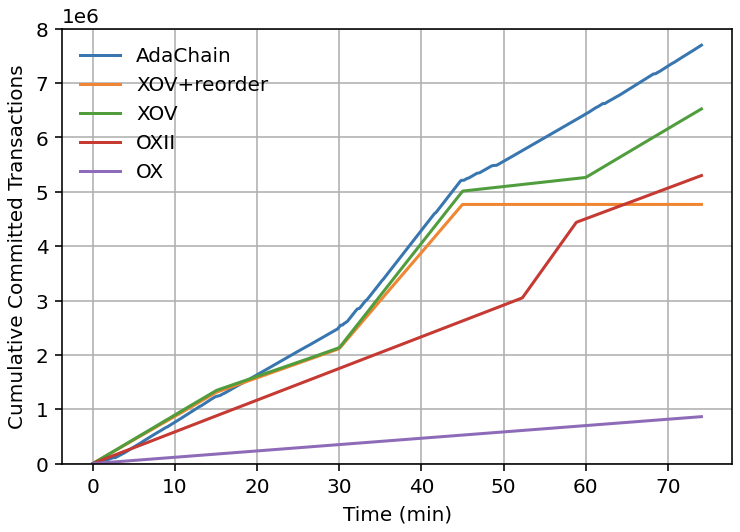}
\vspace{-1em}
\caption{Cumulative committed transactions with time for a changing workload, showing \sys's ability to maintain superior performance during workload shifts.}
\label{fig:changing_cdf}
\end{figure}

Our next experiment focuses on the key benefits of \sys:
\emph{when the workload is changing, \sys can commit significantly more transactions than the best baseline during the same deployment period}.
To emulate a changing workload, we run workload \A for the first $15$ minutes,
followed by workloads \B, \C, \D, and \A, each for 15 minutes.
We use the same four baselines as in Figure~\ref{fig:static}, which are the optimal architectures for workload \A-\D when they are static. 

Figure~\ref{fig:changing_cdf} shows the number of cumulative committed transactions with respect to time.
During the entire $75$ minutes, \sys successfully completed $7.73 \times 10^6$ committed transactions,
while the best baseline XOV completed $6.60 \times 10^6$ committed transactions.
The worst baseline OX only completed $0.87 \times 10^6$ committed transactions.
\sys can successfully commit $1.12$ million ($17\%$) more transactions than the best baseline during $75$ minutes.
The trend in Figure~\ref{fig:changing_cdf} also suggests
the improvement of \sys would become increasingly significant with a longer deployment time
and more variations in the workloads, which are common in today's BaaS. 

Interestingly, Figure~\ref{fig:changing_cdf} also shows a ``catastrophic" effect
for certain fixed architectures when the workload is changing.
For instance, when transitioning back to workload \A again ($60$-$75$ min),
the slope of XOV+reorder is near zero, indicating poor performance where few if any transactions are completed successfully.
However, if we start running XOV+reorder right from the beginning under workload \A without any changes to the workload (Figure~\ref{fig:static}(a)),
XOV+reorder would be the optimal architecture.
XOV+reorder performs poorly in workload \D (45-60 min)
due to the high overhead of Johnson's algorithm with a large number of cycles,
which slows down the block formation.
Since the block formation is sequential, the number of pending blocks grows significantly.
Thus, incoming transactions simulate on stale data and would fail in the MVCC validation phase,
even when transitioning back to workload \A again.
OX suffers from similar problems due to a large number of pending blocks.
This phenomenon also justifies our watermark-based design of \sys, as elaborated in Section~\ref{sec:switching}.

\begin{figure*}
\begin{minipage}{\textwidth}
	\centering
	\includegraphics[width=0.7\linewidth]{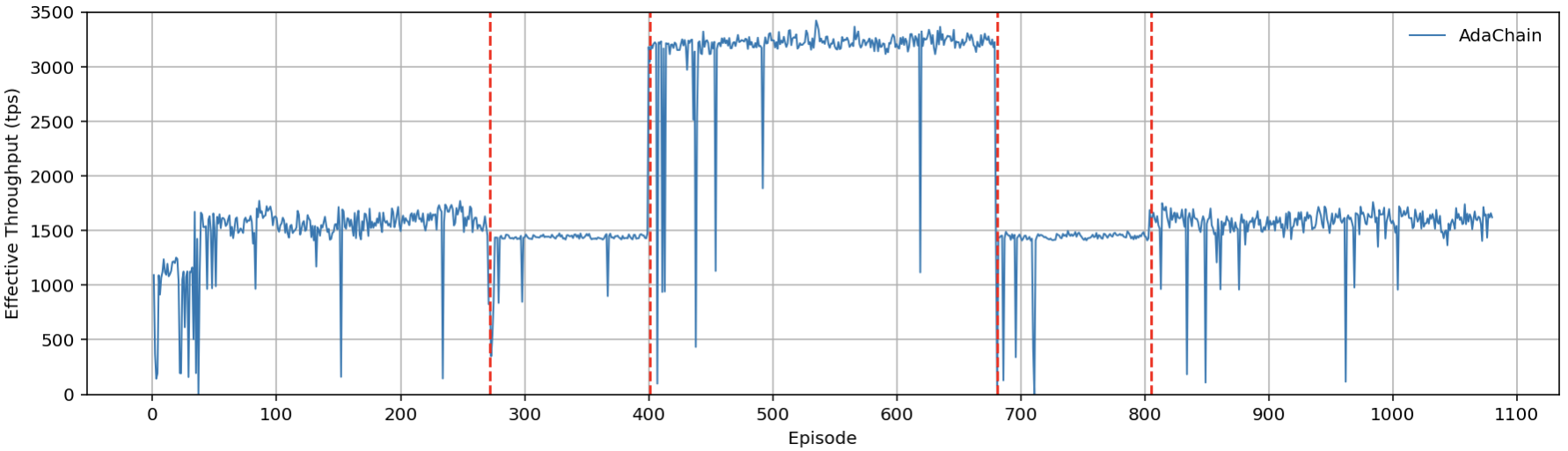}
\end{minipage}
\vspace{-1em}
\caption{Effective throughput of \sys in each episode. Here, red vertical line indicates when the workload shifts. The workload shifts every 15 minutes. The number of episodes per 15 minutes duration varies depending on the transactions arrival rate and compute intensity of workload.}
\label{fig:changing_pdf}
\end{figure*}

To further investigate how \sys switches its architecture under a changing workload,
we plot \sys's effective throughput in each episode during the $75$ minutes in Figure~\ref{fig:changing_pdf}.
The red dashed vertical line indicates when the workload shifts.
Although workloads \A and \B have the same duration in terms of wall clock time ($15$ min),
they vary in terms of the number of episodes.
This is because, depending on the transaction arrival rate and compute intensity of different workloads,
each episode (which is marked by the high watermark) may have a different time duration. 

Figure~\ref{fig:changing_pdf} shows that when workloads shifts,
\sys is able to quickly converge and perform competitively with the optimal architecture.
For instance, when transitioning from workload \A to \B, \sys quickly converges to the new optimal (i.e., OXII)
and achieves a $1450$ tps throughput.
In contrast, while XOV+reorder is optimal under workload \A, as shown in Figure~\ref{fig:static}(b),
it is able to reach only $900$ tps when processing workload \B, even in the best-case scenario where
the catastrophic effect is avoided by starting with workload \B and XOV+reorder right at the beginning.
When transitioning from workload \B to \C, \sys quickly converges to the new optimal (i.e., XOV) and
achieves a $3250$ tps throughput.
In comparison, OXII, which is optimal under workload \B, is able to achieve only $980$ tps under workload \C~ (Figure~\ref{fig:static}(c)).

Due to Thompson sampling, \sys maintains some degree of exploration in the architecture space even after convergence,
so as to avoid getting stuck at the local optimum.
As \sys gains more experiences (i.e., data points) on a certain workload,
the extent of exploration decreases, which is indicated by the less frequent drops within each $15$ minutes period.
Also, when \sys encounters a workload it has seen before (e.g., transition to workload \A again in the last $15$ minutes),
\sys converges much faster than the first time and has less variation in performance.

\new{Ideally, since we use the state of the previous episode to approximate the state of the next episode, \sys can adapt to workloads that shift less frequently than every two episodes ($20$s at most). In practice, due to the exploration performed by CMAB, as long as the workload changes are slower than the convergence time (as shown in Table~\ref{tbl:avg_tp}), AdaChain can still operate effectively.}

\begin{figure*}
\begin{minipage}{0.32\textwidth}
	\centering
	\includegraphics[width=\columnwidth]{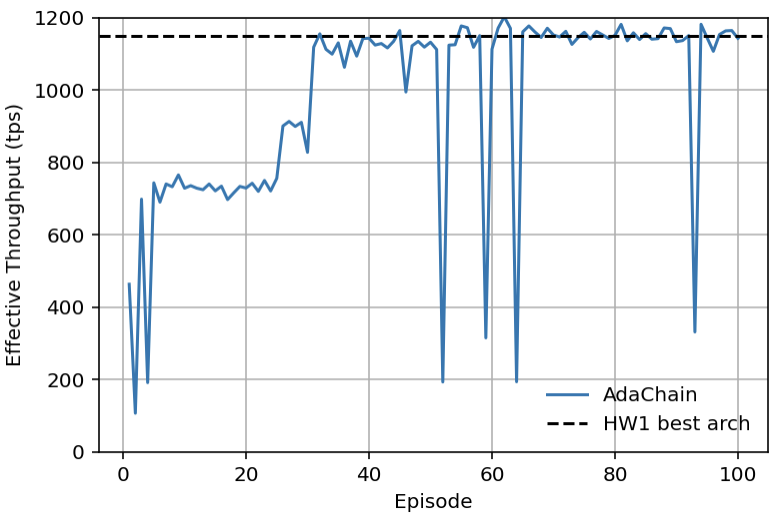}
    \vspace{-2.2em}
    \captionsetup{format=plain, font=scriptsize}
    \caption*{\normalfont \,\,\,\,\,\,\, (a) HW1 (16 cores, RTT=0.15ms, BW=10Gbps)}
\end{minipage}
\begin{minipage}{0.32\textwidth}
	\centering
	\includegraphics[width=\columnwidth]{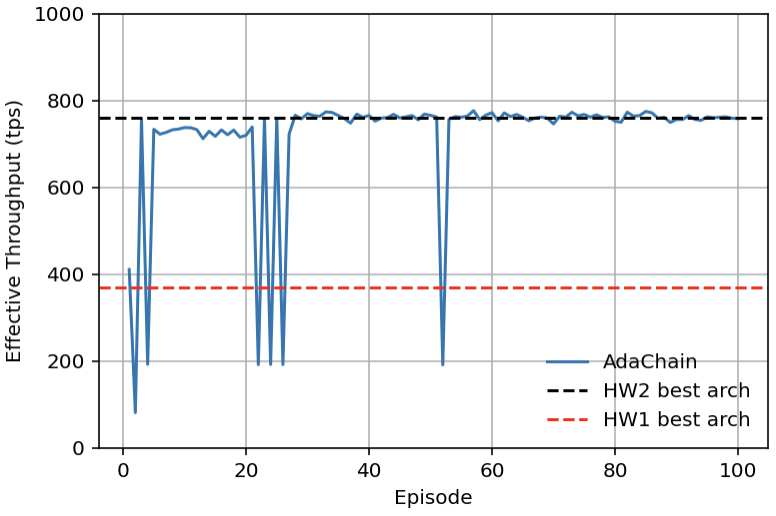}
    \vspace{-2.2em}
    \captionsetup{format=plain, font=scriptsize}
    \caption*{\normalfont \,\,\,\,\,\,\, (b) HW2 (2 cores, RTT=0.15ms, BW=10Gbps)}
\end{minipage}
\begin{minipage}{0.32\textwidth}
	\centering
	\includegraphics[width=\columnwidth]{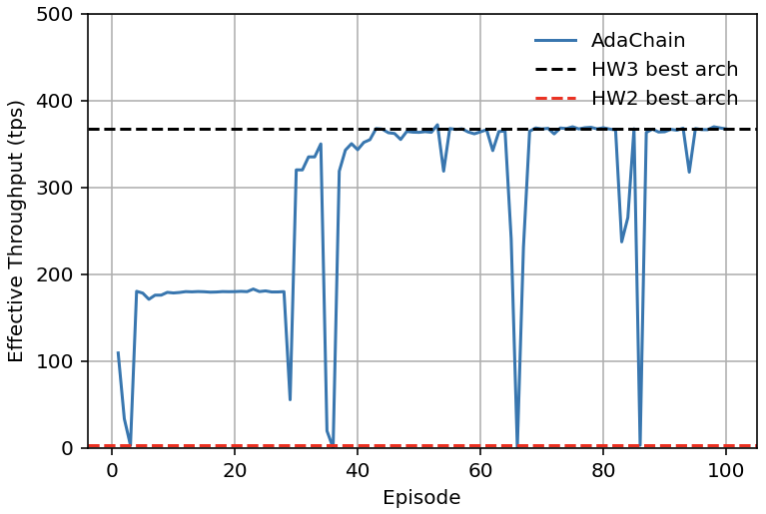}
    \vspace{-2.2em}
    \captionsetup{format=plain, font=scriptsize}
    \caption*{\normalfont \,\,\,\,\,\,\, (c) HW3 (2 cores, RTT=50ms, BW=1Gbps)}
\end{minipage}
\vspace{-0.5em}
\caption{Convergence of \sys to the optimal architecture under different hardware.}
\label{fig:hardware}
\end{figure*}

\subsection{Adaptivity under Different Hardware}\label{sec:exphardware}

Our next set of experiments demonstrates another operational benefit of \sys:
\textit{when deployed on different hardware configurations,
\sys can rapidly converge to the optimal architecture for that hardware
without manually re-configuring the blockchain architecture}.
We use workload \E on three different hardware setups:
HW1 stands for single data center deployment, where the network connecting peers has low latency ($0.15$ ms)
and high bandwidth ($10$ Gbps), and each peer has $16$ CPU cores;
HW2 also stands for single data center deployment, but with $2$ CPU cores per peer;
HW3 stands for a multi-data center deployment, with high latency ($50$ ms) and low bandwidth ($1$ Gbps) network,
and $2$ CPU cores per peer.
As mentioned in Section~\ref{sec:expsetup}, for HW1 and HW2, we use the commodity network fabric;
for HW3, we use the control network as well as Linux {\tt netem}~\cite{netem} to inject delay to the NIC.
For each hardware setup, we also perform a grid search to find the optimal architecture for that hardware:
for HW1, the optimal is (OXII, blocksize = $100$);
for HW2, the optimal is (XOV, blocksize = $1$),
for HW3 the optimal is (OXII, blocksize = $100$) again.
Figure~\ref{fig:hardware} plots \sys's performance in each episode on HW1-HW3,
along with the averaged performance of the optimal architecture for comparison.

As shown in Figure~\ref{fig:hardware}, even under the same workload,
the hardware setup affects the effective throughput and
thus affects the choice of the best architecture.
For instance, OXII performs well when each server has enough compute resources
(HW1 best arch in Figure~\ref{fig:hardware}(a)),
but suffers when the servers have low compute resources
(HW1 best arch in Figure~\ref{fig:hardware}(b));
StreamChain can perform well in a single data center deployment
(HW2 best arch in Figure~\ref{fig:hardware}(b)),
but suffers from the high round trip time when deployed across multiple data centers
due to the small batch size it uses in the consensus protocol
(HW2 best arch in Figure~\ref{fig:hardware}(c)).
No matter what type of hardware \sys is deployed on,
\sys can adapt itself to the optimal architecture for that hardware. 

More importantly, for any kind of unseen hardware setup,
users of \sys do not need to recollect data and retrain the machine learning model offline.
\sys is an online system that learns from its past experiences and balances exploitation and exploration.
With \sys, BaaS can have humans completely out of the loop.

\subsection{Overhead of Learning}\label{sec:expoverhead}

\begin{table}
\caption{Overhead of each stage in \sys.}
\scriptsize
\begin{tabular}{@{}p{0.7cm}cccccc}
\toprule
 &   featurization   & 
  communication  &
 training   & 
 inference  &
 episode  \\ \midrule

mean  & $0.11s \pm 0.01$ & $0.14s \pm 0.04$ & $0.21s \pm 0.04$ & $0.01s \pm 0.01$ & $3.67s \pm 2.12$  \\

median & 0.11s & 0.14s & 0.21s & 0.01s & 2.57s \\ 

max & 0.20s & 0.27s & 0.32s & 0.02s & 16.35s  \\ 
\bottomrule
\end{tabular}

\vspace{0.5em}
\label{tbl:overhead}
\end{table}

Our last experiment evaluates the additional overhead incurred by \sys's learning framework.
We repeat the experiment in Section~\ref{sec:expchanging} and profile every stage that involves the learning agent.
We report the results along with the episode duration in Table~\ref{tbl:overhead}.

Before deriving the architecture for the next episode,
\sys needs to go through feature extraction, communication, training and inference phases in sequence (Section~\ref{sec:overview}).
Table~\ref{tbl:overhead} shows that all of these four phases have low average overhead with low variance,
as compared to the episode duration.
The additional mean overhead time ($0.47$s) is only $12.8\%$ of the average episode duration ($3.67$s). 

More importantly, the $12.8\%$ overhead can be masked by parallelizing transaction processing and learning.
The learning phase only occurs between the low and high watermark period, which constitutes $25\%$ duration with each episode.
Considering that the median episode duration is $2.57$s,
this interval time is $0.64$s which is higher than the mean overhead of $0.47$s.
During this interval, the peers continue to process transactions based on the current architecture,
while in parallel, the learning agent goes through the four stages to determine the architecture for the next episode.
Such parallel execution ensures that the overhead of \sys does not adversely affect its effective throughput,
as long as there are some spare CPU cycles devoted to the learning agent.

Unlike deep neural networks, which are especially expensive to train,
the random forest model used by \sys has moderate training overhead.
With thousands of data points in the experience buffer, \sys only incurs a maximum training overhead of $0.32$ seconds.
Moreover, the succinct action space design also results in a lightweight inference phase,
i.e., $0.02$ seconds.
When \sys is deployed for a long run,
techniques such as periodic resampling and limiting the length of experience buffer~\cite{bao} can be utilized.

\section{Related Work} \label{sec:related}

\sparagraph{Learned systems.}
More generally, many recent works have applied machine learning concepts to systems components. These works, falling under the umbrella of machine programming~\cite{pillars}, cannot be exhaustively enumerated here, but we refer the reader to past work on indexing~\cite{ml_index}, cardinality estimation~\cite{deep_card_est, deep_card_est2}, index selection~\cite{msr_paper}, database tuning~\cite{selfdrivingcidr}, scheduling~\cite{decima}, garbage collection~\cite{learned_gc}, and concurrency control for in-memory databases~\cite{wang2021polyjuice}. \new {As a novel application, learned permissioned blockchains not only require unique featurization of the blockchain design landscape (i.e., action space), but also operate in an environment where there are Byzantine failures. Consequently, our design uses a fully decentralized machine learning approach to handle the untrustworthiness of participating nodes.}

\new {\sparagraph{Database migration.} Efficient and live migration of databases has been studied in multi-tenant data infrastructures~\cite{elmore2011zephyr, das2011albatross, kang2022remus, lin2019mgcrab, cao2022polardb}. Unlike existing work that mainly migrates data between different physical nodes, \sys switches between system architectures within the same participating node. Moreover, \sys's migration protocol is robust to Byzantine failures and mitigates the explorations performed by reinforcement learning.
}

\section{Conclusion}

In this paper, we presented \sys, an adaptive blockchain framework that leverages reinforcement learning to
dynamically switch between different blockchain architectures based on the workload.
\sys is able to identify the optimal blockchain architecture as workload changes, obtaining significantly higher throughput compared to fixed architectures.
As future work, we are exploring expanding the learning framework to cover other aspects of blockchain architectures,
e.g., choosing the best performing consensus protocol.
Another intriguing future direction is to figure out whether our learning framework can be used to uncover new effective architectures not previously explored by human experts.

\begin{acks}
We thank the anonymous reviewers for their insightful feedback and suggestions.
This work is funded by NSF grants CNS-2104882, and CNS-1703936 and by NSF/Intel joint grant \#2011168.
\end{acks}

\balance

\bibliographystyle{ACM-Reference-Format}
\bibliography{_blockchain,_privacy,_system,ryan-cites-long}

\end{document}